\documentclass[12pt]{article}

\usepackage{newtxtext,newtxmath}

\usepackage{graphicx}

\usepackage[letterpaper,margin=1in]{geometry}

\renewenvironment{abstract}
	{\quotation}
	{\endquotation}

\date{}

\makeatletter
\renewcommand{\fnum@figure}{\textbf{Figure \thefigure}}
\renewcommand{\fnum@table}{\textbf{Table \thetable}}
\makeatother

\usepackage{scicite}

\usepackage{url}

\usepackage[normalem]{ulem}
\usepackage{xcolor}

\def\scititle{
	Understanding the Density Maximum of Water with Machine Learned Potentials
}
\title{\bfseries \boldmath \scititle}

\author{
	Yizhi Song$^{1}$,
	Renxi Liu$^{1}$,
	Chunyi Zhang$^{2,1}$,
    Yifan Li$^{3}$,
    Biswajit Santra$^{1 \dagger}$,\and
    Mohan Chen$^{4,5}$,
    Michael L. Klein$^{6,7}$,
    Xifan Wu$^{1,6 \ast}$\and
	\small$^{1}$Department of Physics, Temple University, Philadelphia, Pennsylvania 19122, USA.\and
	\small$^{2}$Eastern Institute of Technology, Ningbo, Zhejiang 315200, P. R. China.\and
    \small$^{3}$Department of Chemistry, Princeton University, Princeton, New Jersey 08544, USA.\and
    \small$^{4}$HEDPS, CAPT, College of Engineering and School of Physics, Peking University, Beijing 100871, P. R. China.\and
    \small$^{5}$Academy for Advanced Interdisciplinary Studies, Peking University, Beijing 100871, P. R. China.\and
    \small$^{6}$Institute for Computational Molecular Science, Temple University, Philadelphia, Pennsylvania 19122, USA.\and
    \small$^{7}$Department of Chemistry, Temple University, Philadelphia, Pennsylvania 19122, USA.\and
	\small$^\ast$Corresponding author. Email: xifanwu@temple.edu\and
    \small$^\dagger$Current address: Schrödinger Inc., New York: 1540 Broadway, New York, NY 10036, USA.
}

\begin{document} 

\maketitle

\begin{abstract} \bfseries \boldmath
After melting, at ambient pressure, the density of water continues to increase with temperature until it reaches a maximum around 4 °C. For nearly a century, this phenomenon has been qualitatively attributed to a mixture of ordered and disordered structures. Herein, we employ a deep neural network to train a machine learned (ML) interatomic potential for water using electronic structure data from advanced density functional theory. Notably, molecular dynamics simulations with the ML potential reproduce both the experimental water density anomaly and the thermal expansion coefficient. Detailed structural analysis of the computed hydrogen-bond network reveals that the density anomaly arises from an emergent liquid structure that retains nearly ideal tetrahedral coordination at short range but collapses at intermediate range. Our findings point to a more delicate mechanism causing the density maximum than the conventional picture, emphasizing the collective roles of structural orderings at different length scales.
\end{abstract}


\section*{Introduction}
\noindent
Among water’s many unique properties, its density anomaly is arguably the most significant and consequential~\cite{eisenberg2005}. Unlike typical liquids, water becomes denser upon heating from its melting point, reaching a maximum near 4 °C before decreasing in density with further warming. This nonmonotonic behavior has profound ecological implications. On a more fundamental level, the anomaly challenges conventional liquid-state theories and has inspired new theoretical models for network-forming liquids, pushing the boundaries of classical physics.

The density anomaly remained poorly understood until the seminal work of Bernal and Fowler in 1933~\cite{bernal1933}, which provided the first microscopic structure model. They proposed that water’s anomalous behavior stems from its hydrogen-bond (H-bond) network, which at low temperatures forms an ordered, ice-like structure characterized by large open volumes and low density. As temperature rises, thermal fluctuations disrupt the H-bonds, transforming the network into a more disordered configuration, thereby enabling denser molecular packing. The Bernal–Fowler model laid the conceptual foundation for later developments based on computer simulations and experiments~\cite{poole1992, bellissentfunel1998, huang2009, soper2000, cuthbertson2011, Santra2015, gonzalez2022, mishima1998, Nilsson2015}, which describes liquid water as a dynamic mixture of low-density, tetrahedral-like and high-density, disordered local structures, associating the anomalies with a liquid–liquid critical point~\cite{poole1992, russo2014, mishima1998, Nilsson2015, xu2005, stanley2008, tanaka2012bond}.

Despite these advances, the microscopic origin of density anomaly remains unresolved and actively debated~\cite{gonzalez2022, cho1996, Vega2005, morawietz2016, finney2024}. The hypothesized liquid–liquid critical point is not directly accessible through experiment~\cite{gallo2016two}, necessitating reliance on computational investigations. In statistical theory~\cite{cuthbertson2011, Santra2015, gonzalez2022, russo2014}, free energies are typically modeled using order parameters~\cite{singh2016, Duboue-Dijon2015, Shi2018} that characterize local structures within the H-bond network responsible for density fluctuations, classifying local structures into low-density and high-density states. However, the microscopic features identified and the population fraction assigned to each structural type are highly sensitive to the choice of structural descriptor~\cite{Shi2018, Tanaka2019}. This sensitivity highlights the need for a precise understanding of how water molecules are locally packed within the H-bond network in order to unambiguously link microscopic structure to density fluctuations. 

\textit{Ab initio} molecular dynamics (AIMD)~\cite{car1985} based on density functional theory (DFT)~\cite{hohenberg1964, kohn1965} provides a rigorous quantum mechanical framework for probing water's behavior~\cite{Swartz13, Gaiduk17, Rozsa20, Fujie21, dellostritto2020, Santra2015}, but its high computational cost has traditionally limited its application. Recent advances in machine learning~\cite{behler2007, zhang2018prl, han2018ccp, wang2018cpc} now enable large-scale simulations with DFT-level accuracy~\cite{Hou20, Song2025}. While promising, these models have often predicted temperatures of maximum density (TMD) below the melting point~\cite{morawietz2016, Li2024}, contradicting experiment. This discrepancy underscores the highly delicate nature of H-bond network and the continuing need for advanced theories.

Here, we address the microscopic origin of water's density anomaly using large-scale molecular simulations based on the Deep Potential (DP) framework~\cite{zhang2018prl, han2018ccp, wang2018cpc}. A deep neural network potential, trained on data from a van der Waals (vdW)-inclusive hybrid PBE0 functional~\cite{perdew1996prl, perdew1996jcp, adamo1999, wu2009, tkatchenko2009}, is employed to simulate water.

\section*{Results}

\begin{figure}
\centering
\includegraphics[width=0.85\linewidth]{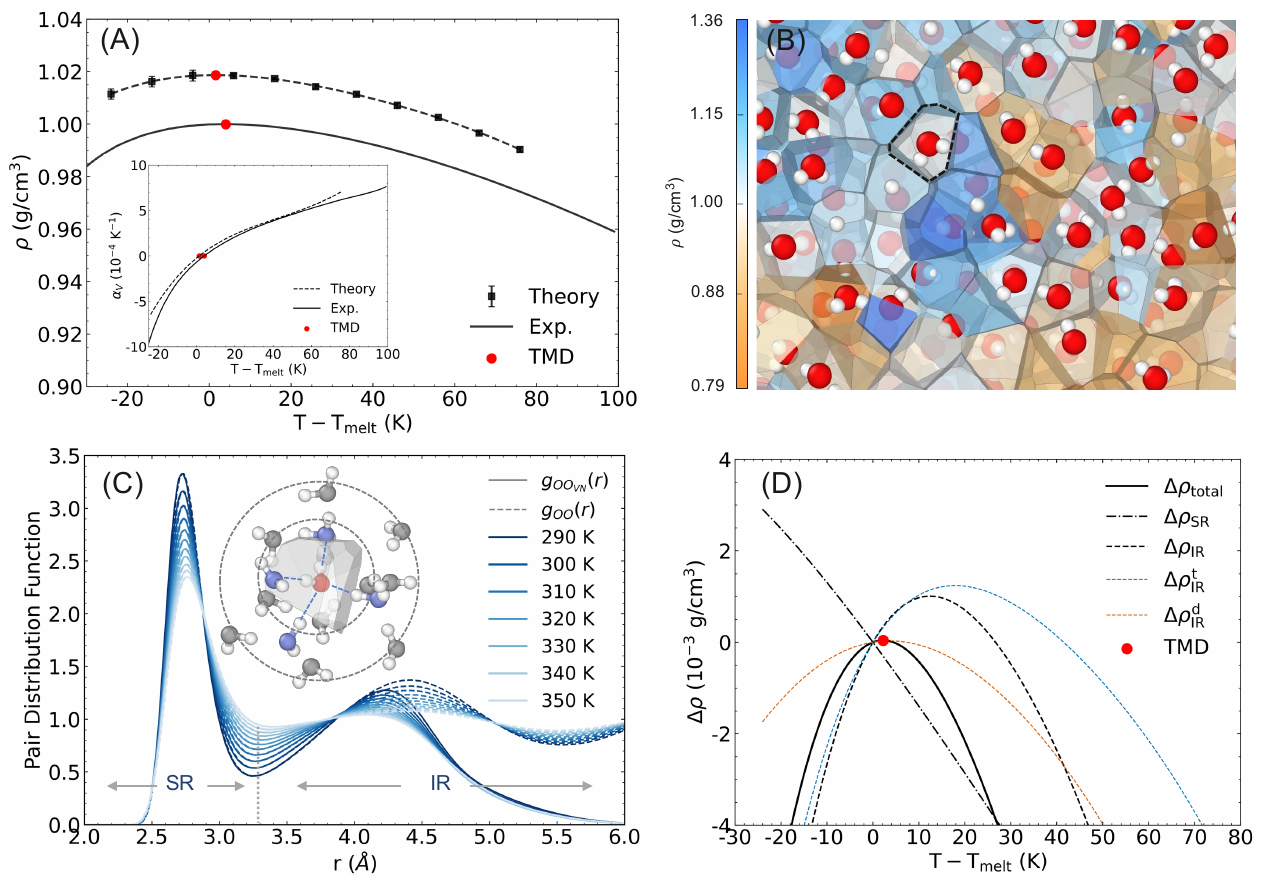}
\caption{\textbf{Structural decomposition of water’s density anomaly.}
(\textbf{A}) Temperature-dependent water density and thermal expansion coefficient (inset) from theory and experiment~\cite{kell1967, hare1987}. Red circles indicate the TMD. All curves are horizontally shifted by the melting temperature (314.0~K).  
(\textbf{B}) Snapshot of liquid water at 330 K (approximately 16~K above the melting point), with Voronoi cells color-coded by local density. The color scale ranges from low local density (orange) to high local density (blue). A representative Voronoi cell is highlighted with a black outline.
(\textbf{C}) Oxygen–oxygen PDFs, \( g_{\mathrm{OO}}(r) \) (dashed lines), and the corresponding partial contributions from Voronoi neighbors, \( g_{\mathrm{OO}_{\mathrm{VN}}}(r) \) (solid lines). Colors indicate different simulation temperatures. Inset shows a representative Voronoi cell (gray polyhedron) of a central water molecule (red); neighbors forming H bonds with the center molecule are marked in blue, others in gray. The two gray circles indicate the first and second coordination shells, respectively.
(\textbf{D}) Decomposition of the total density change \( \Delta \rho_{\mathrm{total}} \) into short-range (\( \Delta \rho_{\mathrm{SR}} \)) and intermediate-range (\( \Delta \rho_{\mathrm{IR}} \)) contributions. The intermediate-range component is further divided into contributions from tetrahedral (\( \Delta \rho_{\mathrm{IR}}^{\mathrm{t}} \)) and disrupted tetrahedral (\( \Delta \rho_{\mathrm{IR}}^{\mathrm{d}} \)) structures, as defined in Eq.~(3).
}
\label{fig:Fig1}
\end{figure}

As shown in Fig.~\ref{fig:Fig1}(A), we present the predicted water density as a function of temperature relative to its melting point, with experimental data~\cite{kell1967, hare1987} included. A clear non-monotonic behavior is observed, and the curvature of the simulated density–temperature curve closely follows experimental results across both the supercooled region and temperatures above the melting point. This accurately captured nonlinearity is further supported by the isobaric thermal expansion coefficient $\alpha_v$ shown in the inset of Fig.~\ref{fig:Fig1}(A), which exhibits excellent agreement with experiments~\cite{kell1967, hare1987}. In both the simulated density and thermal expansion curves, the density anomaly is identified at TMD of 1.5~K above the melting point =(314.0~K), with a corresponding maximum density $\rho_{\mathrm{max}}$ = 1.02~g/cm³. The corresponding experimental reference values of the TMD and maximum density are 3.98~°C and 1.00~g/cm³, respectively \cite{kell1967}. By contrast, other exchange–correlation functionals, such as revPBE0-D3, tend to overestimate the difference between TMD and melting point ($\approx$10 K)~\cite{Cheng2019}, while BLYP-vdW incorrectly predicts their relative order~\cite{morawietz2016}. 

\subsection*{Decomposition of local density changes}

Macroscopically, the observed density is an ensemble average over local fluctuations driven by thermal motion. To resolve these, we employ the Voronoi method~\cite{voronoi1908, rycroft2009}, which partitions the water into space-filling polyhedra centered on the oxygen atoms. The boundaries of each cell are defined by perpendicular bisecting planes between a given central oxygen and its Voronoi neighbors. This geometric construction assigns to each water molecule \( i \) a unique local volume \( V_i \), from which the local density is calculated as $\rho_i = m/V_i$, where $m$ is the molecular mass of a water molecule. The macroscopic density is then given by $\rho = m/\langle V \rangle$, where \( \langle V \rangle \) is the ensemble-averaged local volume~\cite{lazar2022}. Accordingly, variations in the macroscopic density are related to the mean local volume as $\delta \rho = -\frac{m}{\langle V \rangle^2} \delta {\langle V \rangle}$. Fig.~\ref{fig:Fig1}(B) shows a representative Voronoi tessellation and the corresponding distribution of $\rho_i$ from an equilibrated snapshot at 330~K. As expected, pronounced spatial inhomogeneity is observed, with broad fluctuations in $\rho_i$.

Furthermore, the local density $\rho_i$ of a given water molecule depends crucially on the spatial distribution of its Voronoi neighbors. In Fig.~\ref{fig:Fig1}(C), we present the pair distribution function (PDF) between a central oxygen atom and the oxygen atoms of its Voronoi neighbors, alongside the overall oxygen-oxygen PDF for comparison. The Voronoi neighbors consist of all molecules from the first coordination shell and the interstitial region, with only partial contributions from the second coordination shell in Fig.~\ref{fig:Fig1}(C). As the distance from the central molecule increases, the number of Voronoi neighbors decays rapidly and vanishes near $\sim$ 6~\AA\, consistent with the theoretical density-density correlation length~\cite{English2011}.

The Voronoi-based analysis above indicates that the local density is determined by the packing of water molecules in the H-bond network at both short range (SR) and intermediate range (IR). Compared to the tetrahedral structure formed by intact H-bonds in ice, the H-bond network in water is highly dynamic. Under thermal fluctuations, H-bonds continuously break and reform on the picosecond timescale~\cite{laage2006molecular, laage2012water}, yielding a partially collapsed H-bond network. At SR, the structure of the H-bond network can be described by either a near-tetrahedral structure with four H-bonds (see inset of Fig.~\ref{fig:Fig1}(C)), or a disrupted tetrahedron with one or more broken H-bonds, centered on a water molecule. Due to these clear structural characteristics at SR, each water molecule in the equilibrated trajectory can be unambiguously assigned to either a tetrahedral or a disrupted tetrahedral structure, with corresponding Voronoi cell volumes $\langle V_\mathrm{t} \rangle$ and $\langle V_\mathrm{d} \rangle$, and local densities \( \langle \rho_\mathrm{t} \rangle \) and \( \langle \rho_\mathrm{d} \rangle \), respectively. Accordingly, the ensemble-averaged local volume $\langle V \rangle$ can be rigorously decomposed as:
\begin{align}
\langle V \rangle = n_\mathrm{t} \langle V_\mathrm{t} \rangle + (1 - n_\mathrm{t}) \langle V_\mathrm{d} \rangle,
\end{align}
where \( n_\mathrm{t} \) and \( 1-n_\mathrm{t} \) denote the fractions of water molecules adopting tetrahedral and disrupted tetrahedral structures, respectively.

To separate the contributions from SR and IR ordering in the H-bond network to the overall density variation, we apply a constrained variational decomposition method to the macroscopic density \( \rho \) as follows: 
\begin{align}
\delta \rho 
    &= A \Bigg[ 
    \underbrace{\left( \langle \rho_\mathrm{t} \rangle - \langle \rho_\mathrm{d} \rangle \right) \delta n_\mathrm{t}}_{\text{SR}} + \underbrace{n_\mathrm{t} k \, \delta \langle \rho_\mathrm{t} \rangle + (1 - n_\mathrm{t}) \frac{1}{k} \, \delta \langle \rho_\mathrm{d} \rangle}_{\text{IR}} 
\Bigg] \nonumber \\
&= \delta \rho_{\text{SR}} 
    + \delta \rho_{\text{IR}}^{\mathrm{t}} 
    + \delta \rho_{\text{IR}}^{\mathrm{d}}.
\end{align} 
Here, \( \delta \rho_{\text{SR}} \) quantifies the contribution from changes in the population fraction of tetrahedral structures, \( \delta n_\mathrm{t} \), which is governed solely by SR H-bond ordering. The terms \( \delta \rho_{\text{IR}}^{\mathrm{t}} \) and \( \delta \rho_{\text{IR}}^{\mathrm{d}} \) denote the IR contributions to the density change in tetrahedral and disrupted tetrahedral structures, respectively. By holding the SR ordering fixed within each structural type, the remaining variations in local density, $\delta \langle \rho_\mathrm{t} \rangle$ and $\delta \langle \rho_\mathrm{d} \rangle$, can be attributed entirely to changes in IR ordering, giving rise to \( \delta \rho_{\text{IR}}^{\mathrm{t}} \) and \( \delta \rho_{\text{IR}}^{\mathrm{d}} \). The prefactors are defined as \( A = \rho^2 / \langle \rho_\mathrm{t} \rangle \langle \rho_\mathrm{d} \rangle \) and \( k = \langle \rho_\mathrm{d} \rangle / \langle \rho_\mathrm{t} \rangle \), both of which vary weakly with temperature and can be treated as constants over the studied temperature range (see Supplementary Text, Fig.~\ref{fig:prefactors}). The total cumulative change in density with respect to temperature can be obtained by integrating Eq.~(2):
\begin{align}
\Delta \rho_{\mathrm{total}} 
&= \int_{T_{\mathrm{melt}}}^{T} \delta \rho(T')\, dT'
= \Delta \rho_{\mathrm{SR}} 
+ \Delta \rho_{\mathrm{IR}}^{\mathrm{t}} 
+ \Delta \rho_{\mathrm{IR}}^{\mathrm{d}}.
\end{align} 
The calculated contributions are shown in Fig.~\ref{fig:Fig1}(D).

\subsection*{Density variation from short-range ordering in the H-bond network}
\begin{figure}
\centering
\includegraphics[width=0.5\linewidth]{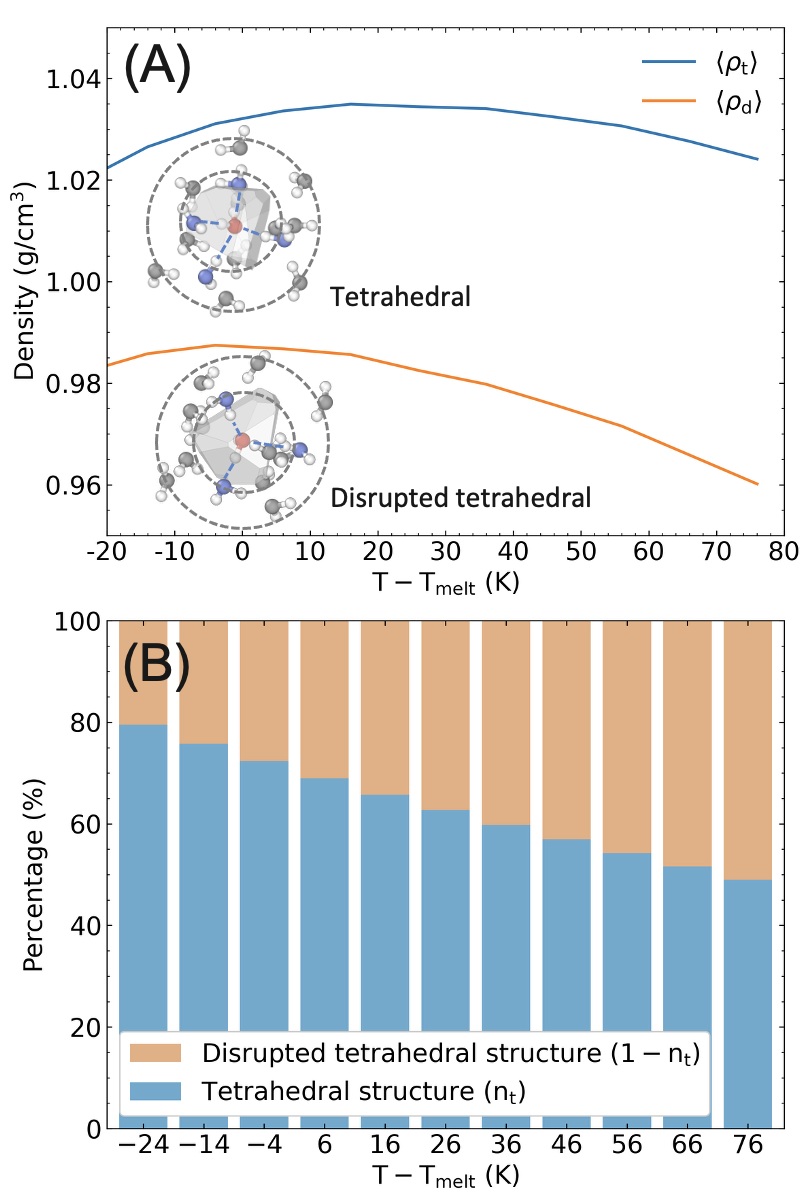}
\caption{\textbf{Short-range ordering in the H-bond network.}
(\textbf{A}) Temperature-dependent local densities of tetrahedral (\( \langle \rho_t \rangle \)) and disrupted tetrahedral (\( \langle \rho_d \rangle \)) structures. Insets show the representative molecular structures for each structural type. The disrupted tetrahedral structure exhibits a larger Voronoi cell volume.
(\textbf{B}) Temperature-dependent population fractions of tetrahedral (\( n_\mathrm{t} \)) and disrupted tetrahedral structures (\( 1-n_\mathrm{t} \)).
}
\label{fig:sr}
\end{figure}

The intrinsic density fluctuations in the H-bond network is driven by the dynamic and directional nature of hydrogen bonding, which continuously interconverts water molecules between tetrahedral structures with local density $\langle \rho_\mathrm{t} \rangle$ and disrupted tetrahedral structures with local density $\langle \rho_\mathrm{d} \rangle$, through the breaking and reformation of H-bonds. At temperatures below TMD, SR ordering in water is dominated by near-tetrahedral arrangements. The Voronoi cell associated with this locally ordered structure forms a compact polyhedron of volume $\langle V_\mathrm{t} \rangle$ surrounding the central molecule, as schematically illustrated in the inset of Fig.~\ref{fig:sr}(A). At elevated temperatures, thermal fluctuation weakens and breaks H-bonds, disrupting the SR H-bond network. As the number of broken H-bonds grows, molecules that were previously H-bonded to the central molecule are displaced from the first coordination shell into interstitial regions as non-bonded neighbors, turning into disrupted tetrahedral arrangements. The associated Voronoi cell expands to a larger volume $\langle V_\mathrm{d} \rangle$ with increased open space, leading to a lower local density $\langle \rho_\mathrm{d} \rangle$, as shown in Fig.~\ref{fig:sr}(A). It should be noted that the local density $\langle \rho_\mathrm{t} \rangle$ is consistently higher than $\langle \rho_\mathrm{d} \rangle$, resulting in a positive gap between $\langle \rho_\mathrm{t} \rangle$ and $\langle \rho_\mathrm{d} \rangle$ across the entire temperature range studied. This is expected, as local density is primarily influenced by the packing of nearest neighbors, governed by the SR ordering.

With increasing temperature, the fraction \( n_\mathrm{t} \) of water molecules exhibiting near-tetrahedral local ordering (and higher local density $\langle \rho_\mathrm{t} \rangle$) decreases, while the fraction \( 1-n_\mathrm{t} \) corresponding to disrupted tetrahedral structures (and lower local density $\langle \rho_\mathrm{d} \rangle$) increases, as shown in Fig.~\ref{fig:sr}(B). This structural shift thus results in a monotonic decrease in the density variations from changes in SR ordering $\Delta \rho_{\text{SR}}$, as shown in Fig.~\ref{fig:Fig1}(D). The sharp decline reflects substantial changes in molecular packing within the H-bond network at SR, underscoring the important role of SR ordering in governing water's density anomaly. Furthermore, the approximately linearly decreasing trend of $\Delta \rho_{\text{SR}}$ with temperature is also anticipated, given that the characteristic energy of hydrogen bonding is roughly two orders of magnitude greater than the thermal energy over the temperature range considered, placing the system firmly within the linear response regime. Nevertheless, despite its significant magnitude, the density change arising from SR H-bond ordering alone cannot account for the emergence of TMD, which requires a non-monotonic density variation with respect to temperature.

\subsection*{Density variation from intermediate-range ordering in the H-bond network}
With the SR ordering constrained as tetrahedron in $\Delta \rho_{\mathrm{IR}}^{\mathrm{t}}$ and disrupted tetrahedron in $\Delta \rho_{\mathrm{IR}}^{\mathrm{d}}$, respectively, $\Delta \rho_{\mathrm{IR}} = \Delta \rho_{\mathrm{IR}}^{\mathrm{t}} + \Delta \rho_{\mathrm{IR}}^{\mathrm{d}}$ can be unambiguously attributed to density variations arising from the IR ordering in the H-bond network. In contrast to the nearly linear decrease in $\Delta \rho_{\text{SR}}$, the IR contribution $\Delta \rho_{\text{IR}}$ exhibits a pronounced turnover as a function of temperature, as shown in Fig.~1(D). This non-monotonic behavior is crucial for the emergence of TMD in water. 

A closer examination in Fig.~1(D) further reveals that the turnover effect in $\Delta \rho_{\mathrm{IR}}$ is mainly driven by the component of $\Delta \rho_{\mathrm{IR}}^{\mathrm{t}}$ with strong non-monotonic curvature. It indicates that the density maximum mainly originates from H-bond structural change at IR, however, with near-ideal tetrahedron at SR. The density variation in $\Delta \rho_{\mathrm{IR}}^{\mathrm{t}}$ is contributed by two effects, $\Delta \rho_{\mathrm{IR}}^{\mathrm{t}} = \Delta \rho_{\mathrm{IR}}^{\mathrm{t}\,^{(-)}} + \Delta \rho_{\mathrm{IR}}^{\mathrm{t}\,^{(+)}}$, which influence density in opposite directions, as illustrated in Fig.~3(B) (see the \textit{Structural Analysis} section in Methods for computational details). As temperature increases, the density decrement associated with $\Delta \rho_{\mathrm{IR}}^{\mathrm{t}\,^{(-)}}$ is attributed to the softening of hydrogen bonds. This geometric softening leads to an increased H-bond length $d_{\mathrm{HO\cdots H}}$ between neighboring water molecules, as evidenced by a shift in the center-of-mass distribution of H-bonded neighbors to larger separations (see Fig.~3(A)). Notably, H-bond softening perturbs the geometry of the network without breaking its connectivity, thus preserving the SR H-bond ordering. However, it impacts the IR structure, causing a nearly homogeneous expansion of the Voronoi cells of each water molecule, thereby has a negative effect on the local density $\langle \rho_\mathrm{t} \rangle$ and contributes decreasing $\Delta \rho_{\mathrm{IR}}^{\mathrm{t}\,^{(-)}}$, as schematically shown in Fig.~3(B).

\begin{figure}
\centering
\includegraphics[width=0.5\linewidth]{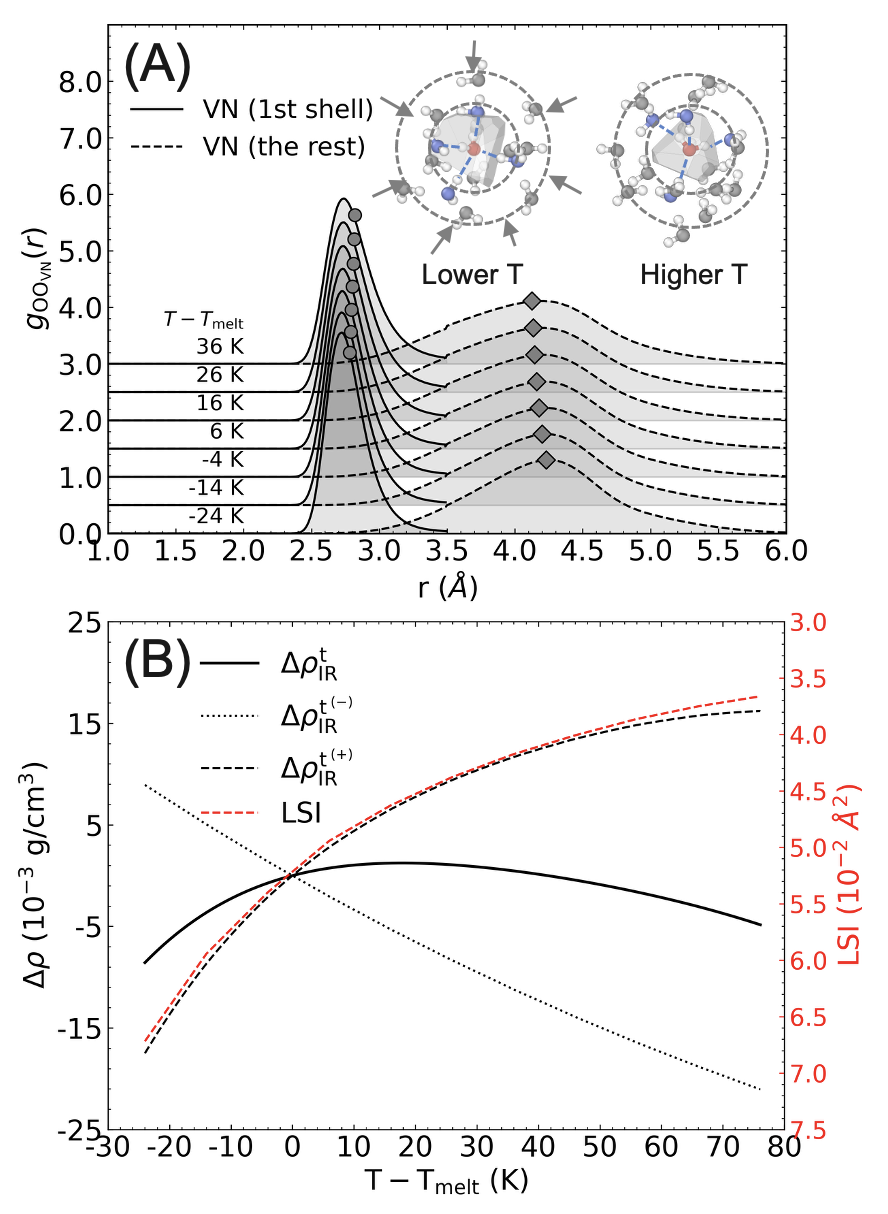}
\caption{\textbf{Intermediate-range ordering in the H-bond network.}
(\textbf{A}) Temperature dependent PDFs between a central water molecule forming a tetrahedral structure and its Voronoi neighbors. Each PDF is decomposed into contributions from Voronoi neighbors in the first coordination shell and those beyond it, with circles and diamonds indicating the average radial positions of each group, respectively. Average radial positions and corresponding standard deviation values for each temperature are provided in Table~\ref{tab:variance}. Insets illustrate how the accumulation of interstitial water molecules (between the two gray circles) leads to a reduction in the Voronoi cell volume of the central molecule at higher temperatures. Arrows indicate the inward movement of water molecules at IR.
(\textbf{B}) Decomposition of IR density change of tetrahedral structures \( \Delta \rho_{\mathrm{IR}}^{\mathrm{t}} \) into \(\Delta \rho_{\mathrm{IR}}^{\mathrm{t}\,^{(-)}}\) and \(\Delta \rho_{\mathrm{IR}}^{\mathrm{t}\,^{(+)}}\). The corresponding LSI is shown on the right axis in red.
}
\label{fig:IR}
\end{figure}

Conversely, the term $\Delta \rho_{\mathrm{IR}}^{\mathrm{t}\,^{(+)}}$ contributes to an increase in density with rising temperature, as shown in Fig.~3(B). This density increase originates from the collapse of the H-bond network at IR. Upon heating water above its melting point, the H-bond network undergoes a fundamental transformation at the IR, due to the highly dynamic nature of hydrogen bonding. In contrast to the near-ideal tetrahedral network in crystalline ice, the second coordination shell in water begins to partially collapse, as evidenced by the diminishing second peak in the oxygen-oxygen PDF $g_{\mathrm{OO}}(r)$ (see Fig.~3(A)). As this collapse progresses, water molecules at IR become non-bonded and leave the second shell, filling the interstitial region. As a result, the mass centers of IR neighbors shift inward toward the central molecule as seen in Fig.~3(A), leading to a decrease in Voronoi cell volumes, as schematically plotted in the inset of Fig.~3(A). As shown in Fig.~3(B), $\Delta \rho_{\mathrm{IR}}^{\mathrm{t}\,^{(+)}}$ rises rapidly at low temperatures, which is attributed to the initially relatively empty interstitial region. This increment slows at higher temperatures, as the population of interstitial water molecules approaches saturation~\cite{soper2000}. This structural transformation is also effectively characterized by the local structure index (LSI)~\cite{Wikfeldt2011}, as shown in Fig.~3(B). Note that the LSI here is computed exclusively for central water molecules adopting a tetrahedral structure, thus has a high correlation with the IR ordering. The temperature dependence of the LSI closely follows the trend of $\Delta \rho_{\mathrm{IR}}^{\mathrm{t}\,^{(+)}}$, providing strong support for the proposed mechanism. 

During the above process, in the softened and increasingly disordered structure, the role of directional H-bonds as the dominant cohesive force diminishes. In turn, vdW interactions, though weaker, become relatively more important in stabilizing the liquid phase. Due to their nondirectional nature, vdW interactions can attract and confine those non-H-bonded water molecules that have moved out of the second shell~\cite{morawietz2016, gaiduk2015}. Here, vdW interactions play a critical role---without them, the density increment in $\Delta \rho_{\mathrm{IR}}^{\mathrm{t}\,^{(+)}}$ would be absent, and the density anomaly would not occur (see Supplementary Text, Fig.~\ref{fig:pbe}).

Comparatively, the turnover behavior in $\Delta \rho_{\mathrm{IR}}^{\mathrm{d}}$ is much weaker, exhibiting a relatively flat profile near the TMD. This indicates that the H-bond network associated with locally disrupted tetrahedral ordering and a collapsed second coordination shell at IR contributes far less to the density maximum of liquid water, making further modifications to the conventional Bernal-Fowler picture. Due to the disrupted SR ordering, the interstitial region becomes populated with non-bonded water molecules originating from the broken tetrahedra. As a result, the density increase due to the collapse of the IR H-bond network is strongly limited by the rapid saturation of interstitial water molecules (see Supplementary Text, Fig.~\ref{fig:disrupted}). Indeed, when $\Delta \rho_{\mathrm{IR}}^{\mathrm{d}}$ is completely neglected, the resulting temperature and density at the maximum density point are barely changed (by 0.15~K and 0.005~g/cm$^3$, respectively).

\section*{Discussion}

By combining machine-learning molecular dynamics with a range-separated analysis of the H-bond network, we provide a microscopic explanation for the density anomaly of water. Previous theoretical efforts have largely focused on identifying local order parameters~\cite{singh2016, Duboue-Dijon2015, Shi2018} that distinguish high- and low-density local environments. While these approaches have provided valuable qualitative insight, they primarily offer structural classification and do not explicitly quantify how H-bond ordering at different length scales contributes cooperatively to density variations. In contrast, our analysis introduces an explicit decomposition of density changes into SR and IR structural contributions. Our results reveal that changes in short-range ordering within the H-bond network alone are insufficient to account for the observed density anomaly; rearrangements in intermediate-range ordering are indispensable. In particular, we find that the non-monotonic density behavior near the TMD arises primarily from a specific structural configuration in the H-bond network---characterized by a collapsed second coordination shell but a nearly ideal tetrahedral arrangement at short range. This configuration yields the most compact molecular packing and thus the greatest density increase near the TMD. This explains why the TMD of water lies just above its melting point, where the majority of water molecules remain tetrahedral, but the H-bond network begins to partially collapse. 

Our analysis also provides insight into the conditions under which the density anomaly vanishes. In aqueous electrolyte solutions, increasing ion concentration progressively disrupts the hydrogen-bond network and reduces the space available for water molecules to form structured coordination shells. We estimate that the density anomaly disappears when the average ion–ion distance becomes too small to form two complete coordination shells around each water molecule, which is an essential condition for the density anomaly to occur. Assuming a uniform distribution of ions, we estimate the critical concentration at which this structural disruption occurs by requiring that the average ion–ion distance equals the characteristic radius encompassing two coordination shells (approximately 4.5~\AA). This yields an estimated threshold concentration of ~2.3~mol/kg for NaCl, consistent with experimental observations, where the TMD vanishes at 2.33~mol/kg~\cite{archer2000}. This agreement supports the proposed mechanism linking intermediate-range hydrogen-bond structuring to the presence of the density anomaly. To our knowledge, such a quantitative prediction of the critical concentration at which the density anomaly disappears has not been provided by previous theoretical frameworks. Overall, this work offers a quantitative and transferable framework for connecting local structure to macroscopic thermodynamic anomalies, with potential applications for other anomalous liquids.

\section*{Materials and Methods}


\subsection*{\textit{Ab Initio} Molecular Dynamics Simulation}

To generate the training data set for deep potential molecular dynamics (DPMD) models, a series of Car-Parrinello~\cite{car1985} \textit{ab initio} molecular dynamics (AIMD) simulations of liquid water were conducted at atmospheric pressure and different temperatures within the isobaric-isothermal (NPT) ensemble in Quantum ESPRESSO~\cite{qe2009}. A hierarchy of different exchange-correlation (XC) functionals with increasing accuracy were adopted, including the semi-local PBE-GGA~\cite{perdew1996prl} functional, the PBE functional with Tkatchenko-Scheffler (TS)~\cite{tkatchenko2009} van der Waals corrections (PBE+vdW), and the hybrid version of the PBE functional~\cite{perdew1996jcp, adamo1999} with TS-vdW corrections (PBE0+vdW) that includes 25\% exact exchange. The simulations were performed with periodically replicated cubic simulation cells containing 128 water molecules, and Parrinello-Rahman barostat~\cite{parrinello1980} constraining the cell fluctuations to be isotropic. The trajectory length of each functional at a specific temperature is listed in Table~\ref{tab:aimd_trajectories}.

The exact exchange in the PBE0+vdW simulations are efficiently computed with Maximally Localized Wannier Function (MLWF) in real space as described in Refs~\cite{wu2009}. To efficiently compute the exact exchange energy at every AIMD step, the required MLWFs were evaluated by minimizing the total spread using second order damped CP dynamics~\cite{tassone1994} and efficient on-the-fly localization of MLWFs~\cite{sharma2003}. During this procedure, we used a fictitious mass of 500 a.u., a damping coefficient of 0.3, and a time step of 4 a.u. Details on the implementation and convergence of exact exchange energy, force, and stress tensor are given here~\cite{ko2020enabling, ko2021enabling}. The key parameters used here to achieve accurate exact exchange energy (within 0.02\% off the fully converged energy~\cite{ko2020enabling}) are: i) the maximum distance between a pair of MLWF centers ($R_\text{pair}$) = 8~Bohr, ii) the radius for solving Poisson equation ($R_\text{PE}$) for self-pair exchange = 6~Bohr and for non-self-pair exchange = 5~Bohr, iii) the radius for multipole expansion ($R_\text{ME}$) for self-pair exchange = 10 Bohr and for non-self-pair exchange = 7~Bohr.

The CP equations of motion for the nuclear and electronic degrees of freedom were integrated using the standard Verlet algorithm and a time step of 2.0 a.u. ($\approx$ 0.05 fs). The ionic temperatures were controlled with Nosé-Hoover chain thermostats~\cite{martyna1992}, each with a chain length of 4 and a frequency of 60~THz. To achieve rapid equipartition of the thermal energy, we employed one Nosé-Hoover chain thermostat per atom (i.e., the so-called “massive” Nosé-Hoover thermostat), and also rescaled the fictitious thermostat masses by the atomic masses, so that the relative rates of the thermostat fluctuations were inversely proportional to the masses of the atoms to which they were coupled~\cite{tobias1993}.

The core electrons were treated with optimized norm-conserving Hamann-Schlüter-Chiang-Vanderbilt (HSCV) pseudopotentials~\cite{hamann1979,vanderbilt1985}, while the valence (pseudo-)wavefunctions were represented explicitly with a planewave basis set. The electronic wavefunctions were expanded using a plane wave basis set with a kinetic energy cutoff of 130~Ry. To ensure an adiabatic separation between the electronic and nuclear degrees of freedom in the CP dynamics, we used a fictitious electronic mass of 100~a.u. and the nuclear mass of deuterium for each hydrogen atom. Mass preconditioning was applied to all Fourier components of the electronic wavefunctions having a kinetic energy greater than 25~Ry~\cite{tassone1994}. To maintain a constant planewave kinetic energy cutoff of 130~Ry during the NpT simulation, we followed the procedure of Bernasconi et al.~\cite{bernasconi1995} by choosing: (i) a cubic reference cell (with L = 30.62 Bohr) that is large enough to cover the fluctuations along each lattice vector of the simulation cell throughout the NpT trajectories, and (ii) a corresponding planewave basis set with a larger kinetic energy cutoff of 150 Ry. During the NpT simulation, planewaves with a kinetic energy beyond the desired cutoff of 130 Ry were smoothly penalised by changing
\begin{align}
G^2 \rightarrow G^2 + A\left[1 + \operatorname{erf}\left(\frac{\frac{1}{2}G^2 - E_0}{\sigma}\right)\right],
\end{align}
with a judicious choice of parameters (A = 200~Ry, $\sigma$ = 15~Ry, $E_0$ = 130~Ry), this modification to $G^2$ causes the higher-energy ($>$ 130 Ry) planewaves to become essentially inactive basis functions in the description of the valence (pseudo-)wavefunctions, and thereby leads to NpT dynamics which mimic a constant planewave cutoff of $\approx$ 130 Ry.

Although a linear-scaling exact-exchange algorithm~\cite{wu2009} was applied to reduce the computational burden of PBE0+vdW calculations, the computation cost is still much higher than PBE and PBE+vdW calculations. Therefore, 48~ps PBE0+vdW AIMD simulations are conducted, which are much shorter than the 216.6~ps PBE trajectory and the 372.3~ps PBE+vdW trajectory as shown in Table~\ref{tab:aimd_trajectories}. As the structure of liquid water predicted by PBE0+vdW are softer than the other two functionals, the 48~ps trajectory can cover most of the potential energy surface and produce a relatively converged deep potential model.

\subsection*{Deep Potential Molecular Dynamics Simulation}

We used the Deep Potential Molecular Dynamics (DeePMD) framework~\cite{zhang2018prl, han2018ccp, wang2018cpc} to perform simulations of liquid water. For each functional, the deep potential model is trained independently using the corresponding AIMD trajectories. The atomic position, total potential energy $E$, ionic forces $\mathbf{F}_i$, and the stress tensor $\mathbf{\Xi}$ at each time step were extracted from the AIMD trajectories and adopted as the input training data for the DPMD model. The training was conducted for $10^6$ steps using the DeePMD-kit package~\cite{wang2018cpc} interfaced with the Tensorflow library~\cite{abadi2016} following the procedure described in Refs.~\cite{zhang2018prl, zhang2018nips}. First, the input data were transformed to local coordinate frames for every atom and its neighbors inside a cutoff distance of 6~Å to preserve the translational, rotational, and permutational symmetries of the environment. This cutoff is sufficiently large to encompass the dominant contributions from hydrogen-bond networks and van der Waals interactions in liquid water, and is consistent with the density--density correlation length of liquid water reported in previous theoretical studies\cite{English2011}. Then the Adam method~\cite{kingma2017} was applied to optimize the deep neural network parameters with the loss function:
\begin{align}
\mathcal{L}(p_\epsilon, p_f, p_\xi) = p_\epsilon \Delta \epsilon^2 + \frac{p_f}{3N} \sum_i |\Delta \mathbf{F}_i|^2 + \frac{p_\xi}{9} \| \Delta \boldsymbol{\xi} \|^2,
\end{align}
where $\Delta \epsilon$, $\Delta \mathbf{F}_i$, and $\Delta \boldsymbol{\xi}$ represent the differences between the training data and current DPMD prediction for the quantities $\epsilon \equiv E/N$, $\mathbf{F}_i$, and $\boldsymbol{\xi} \equiv \mathbf{\Xi}/N$, respectively; $N$ is the number of atoms, and $p_\epsilon$, $p_f$, and $p_\xi$ are tunable prefactors. In the training process, $p_\epsilon$ progressively increases from 0.02 to 1, while $p_f$ progressively decreases from 1000 to 1 for all three DPMD models. The prefactor of the stress, $p_\xi$, progressively increases from 0.02 to 1 for the PBE model, while $p_\xi$ was set to zero in the training progress of PBE+vdW and PBE0+vdW models. The resulting root-mean-square error between the energy predicted by the PBE and PBE+vdW (PBE0+vdW) DPMD models and that predicted by AIMD is smaller than 0.4 (0.8) meV/atom.

The obtained DPMD models were applied to conduct DPMD simulations using the DeePMD-kit package~\cite{wang2018cpc}. For each exchange-correlation functional, a series of DPMD simulations were carried out in the NPT ensemble at atmospheric pressure and temperatures from 290 to 390~K, with 10~K intervals. Considering that PBE does not show a density maximum in 290–390~K, the simulated temperature range for PBE was extended to 270–390~K. The cell size was enlarged to 1024 water molecules and the simulation lasted for 2~ns at each thermodynamic condition with a timestep of 0.5~fs to produce converged densities and structural properties. The first 50~ps trajectories are discarded for equilibration.

To determine the melting temperature of ice ($T_{\text{melt}}$), we employed the thermodynamic integration (TI) method~\cite{frenkel1984, zhang2021} implemented in the Deep Potential Thermodynamic Integration (DPTI) package~\cite{dpti2024}, to yield the absolute Gibbs free energies, \(G_{\mathrm{ice}}\) and \(G_{\mathrm{liq}}\), of ice and liquid water, respectively. The melting temperature was then determined as the temperature at which these two free energies intersect, i.e., when \(G_{\mathrm{ice}} = G_{\mathrm{liq}}\), indicating thermodynamic equilibrium between the solid and liquid phases.

The simulated density–temperature ($\rho$–$T$) data were fitted to a fourth-order polynomial:
\begin{equation}
\rho(T) = a_0 + a_1 T + a_2 T^2 + a_3 T^3 + a_4 T^4.
\end{equation}
The TMD was determined as the temperature at which the fitted polynomial reaches its maximum by evaluating the polynomial on a refined temperature grid (0.01~K resolution). The fitted coefficients are listed in Table~\ref{tab:polyfit_tmd}. Sensitivity tests varying the polynomial order (4th to 6th) indicated a numerical uncertainty of less than 0.2~K, which is small compared to the systematic differences between functionals and does not affect the qualitative trends or mechanistic conclusions presented.

\subsection*{Structural Analysis}

To characterize local structural fluctuations in liquid water, we performed a Voronoi tessellation of the simulation box using the \texttt{voro++} library~\cite{rycroft2009}, assigning a unique spatial volume to each water molecule. Hydrogen bonds (H-bonds) were identified using the geometric criterion introduced by Luzar and Chandler~\cite{luzar1996}, where a hydrogen bond is considered to exist between two water molecules if $R_{\mathrm{OO}} < 3.5$~\AA\ and $\angle \mathrm{H_D{-}O_D{\cdots}O_A} < 30^\circ$. Each water molecule is then classified into one of two structural categories based on its number of H-bonds. Here, water molecules with four or more H-bonds are labeled as adopting a tetrahedral structure, while those with fewer than four H-bonds categorized as having a disrupted tetrahedral structure.

At the intermediate range (IR), increasing temperature leads to geometric softening of H-bonds and an increase in the average H-bond length $d_{\mathrm{HO\cdots H}}$ between neighboring water molecules. This H-bond elongation results in a nearly homogeneous expansion of the Voronoi cells of each water molecule, thereby decreasing the local densities, and contributing negatively to the IR component of the density change. To model this behavior, we consider an idealized tetrahedral configuration in which the central water molecule forms H-bonds with four nearest neighbors arranged in a perfect tetrahedral geometry, and neglects contributions from more distant neighbors. In this simplified picture, the Voronoi cell volume can be estimated as \( V_\mathrm{est} = \sqrt{3}\, a^3 \), where \( a \) is the H-bond length, and the estimated local density \( \rho_\mathrm{est} = m / V_\mathrm{est} \), where \( m \) is the molecular mass. 

To evaluate the validity of this approximation, we compare the actual mean local density \( \langle \rho_\mathrm{t} \rangle \) computed from simulations with the estimated local density \( \rho_\mathrm{est} \) as a function of $a$. As shown in Fig.~\ref{fig:HBlength}, despite the structural complexity of real liquid water, the trend in \( \langle \rho_\mathrm{t} \rangle \) closely follows the trend predicted by the ideal tetrahedral model. This observation indicates that the local density fluctuations in water are strongly correlated with changes in H-bond length. Therefore, the negative contribution to the density variation from H-bond softening in tetrahedral structures can be approximated as:
\begin{align}
\delta \langle \rho_\mathrm{t} \rangle^{(-)} = \delta \rho_\mathrm{est}=\delta\left(\frac{m}{V_\mathrm{est}}\right) 
= \frac{-\sqrt{3} m}{a^4}\, \delta a.
\end{align}
The remaining part of the local density change, attributed to interstitial accumulation, is then estimated as $\delta \langle \rho_\mathrm{t} \rangle^{(+)} = \delta \langle \rho_\mathrm{t} \rangle - \delta \langle \rho_\mathrm{t} \rangle^{(-)}$. Using Eq.~(2) from the main text, the corresponding contributions to the IR component of the density change are computed as:
\begin{align}
\delta \rho_{\mathrm{IR}}^{\mathrm{t}(-)} = A n_t k\, \delta \langle \rho_\mathrm{t} \rangle^{(-)}, \quad
\delta \rho_{\mathrm{IR}}^{\mathrm{t}(+)} = A n_t k\, \delta \langle \rho_\mathrm{t} \rangle^{(+)}.
\end{align}
A similar procedure is applied to evaluate the IR contributions from disrupted tetrahedral structures.

\newpage

\clearpage 

%
\bibliography{science_template} 
\bibliographystyle{sciencemag}

%
%
%
%
%
%


\section*{Acknowledgments}
\paragraph*{Funding:}
This research was supported by the Computational Chemical Center: Chemistry in Solution and at Interfaces, funded by the U.S. Department of Energy (DOE) under Award No. DE-SC0019394. Computational resources were provided by the National Energy Research Scientific Computing Center (NERSC), a DOE Office of Science User Facility, operated under Contract No. DE-AC02-05CH11231. This work was partially supported by National Science Foundation through Grants No. DMR-2053195 and Seven Research, LLC.


\paragraph*{Author contributions:}

Conceptualization: YS, RL, XW

Methodology: YS, RL, CZ, YL, BS

Investigation: YS, RL

Visualization: YS

Supervision: MC, MLK, XW

Writing—original draft: YS, BS, XW

Writing—review \& editing: YS, YL, XW

\paragraph*{Competing interests:}
Authors declare that they have no competing interests.

\paragraph*{Data, Code, and Materials Availability Statement:}
All data needed to evaluate and reproduce the results in the paper are present in the paper and/or the Supplementary Materials. The simulation input files, trained machine-learned potential models, and analysis scripts used in this study are available from the corresponding author upon reasonable request. This study did not generate new materials.

\newpage


\renewcommand{\thefigure}{S\arabic{figure}}
\renewcommand{\thetable}{S\arabic{table}}
\renewcommand{\theequation}{S\arabic{equation}}
\renewcommand{\thepage}{S\arabic{page}}
\setcounter{figure}{0}
\setcounter{table}{0}
\setcounter{equation}{0}
\setcounter{page}{1} 


\begin{center}
\section*{Supplementary Materials for\\ \scititle}

	Yizhi Song,
	Renxi Liu,
	Chunyi Zhang,
    Yifan Li,
    Biswajit Santra,\\
    Mohan Chen,
    Michael L. Klein,
    Xifan Wu$^{\ast}$\\
\small$^\ast$Corresponding author. Email: xifanwu@temple.edu\\
\end{center}

\subsubsection*{This PDF file includes:}
Supplementary Text\\
Figures S1 to S8\\
Tables S1 to S4\\


\newpage


\subsection*{Supplementary Text}
\subsubsection*{Comparison Across Different Exchange-Correlation Functionals}

To evaluate the performance of different XC functionals in describing the thermodynamic properties of water, we compare the melting temperatures ($T_\mathrm{melt}$) and temperatures of maximum density (TMD) obtained from DPMD simulations using the PBE, PBE+vdW, and PBE0+vdW functionals. The results are summarized in Table~\ref{tab:Tmelt_TMD}. The PBE functional significantly overestimates $T_\mathrm{melt}$ and fails to reproduce a density maximum in the studied temperature range, reflecting its known tendency to overstrengthen H-bonds and underestimate structural fluctuations. Including van der Waals corrections (PBE+vdW) improves the predictions considerably: the predicted $T_\mathrm{melt}$ moves closer to the experimental value, and a nonmonotonic density profile emerges. However, the TMD is slightly below the melting point, contrary to experiment. The PBE0+vdW functional further improves the agreement with experiment, predicting $T_\mathrm{melt}$ and TMD in the correct order and with improved quantitative accuracy. These results underscore the importance of both exact exchange and dispersion interactions corrections in capturing the delicate balance of forces---particularly H-bonding and vdW interactions---that govern the anomalous thermodynamic properties of liquid water.

\subsubsection*{Weak Temperature Dependence of Prefactors $A$ and $k$}

The prefactors defined in Eq.~(2) of the main text, \( A = \rho^2 / \langle \rho_t \rangle \langle \rho_d \rangle \) and \( k = \langle \rho_d \rangle / \langle \rho_t \rangle \), exhibit only weak dependence on temperature across the range investigated. As shown in Fig.~\ref{fig:prefactors}, both $A$ and $k$ remain nearly constant, with $A \approx 1.01$ and $k \approx 0.95$, and vary by less than 3\% over the entire temperature range. This weak variation reflects the relatively stable local density contrast between tetrahedral and disrupted tetrahedral structures under the studied thermodynamic conditions. 

For clarity, we emphasize that treating $A$ and $k$ as temperature-independent is introduced only as a simplifying approximation for interpretating the equation, and all analysis presented in the main text (e.g., Fig.~\ref{fig:Fig1}D and Fig.~\ref{fig:IR}B) use the full temperature-dependent forms $A(T)$ and $k(T)$. To quantify the impact of this approximation, we compare the density decomposition obtained using the full temperature-dependent prefactors with that obtained by fixing $A$ and $k$ to their values at the melting temperature ($A \approx 1.02$, $k \approx 0.96$). As shown in Fig.~\ref{fig:prefactor_comparison}, this approximation introduces only minor quantitative differences in $\Delta \rho_{\mathrm{IR}}^{\mathrm{t}}$ and related terms, well within the uncertainty of the analysis and not affecting the qualitative trends or conclusions.

\subsubsection*{Critical Role of van der Waals Interactions}

To evaluate the role of van der Waals (vdW) interactions in IR packing, we compare DPMD simulations conducted using PBE+vdW and PBE functionals, which differ by the presence or absence of explicit dispersion corrections. As shown in Fig.~\ref{fig:pbe}(a, c), the temperature-dependent pair distribution functions (PDFs) between a central tetrahedral water molecule and its Voronoi neighbors reveal significant differences between the two functionals. In simulations using the PBE+vdW functional (Fig.~\ref{fig:pbe}(a)), the second-shell peak of the oxygen–oxygen PDF exhibits a marked inward shift with increasing temperature, along with enhanced population in the interstitial region, resulting in a decrease in Voronoi cell volume. By contrast, in the PBE simulation (Fig.~\ref{fig:pbe}(c)), the second-shell peak position shows minimal temperature dependence, and the population of interstitial molecules is also noticeably reduced. These features indicate a diminished tendency for interstitial water molecules to accumulate in the absence of vdW interactions.

Figure~\ref{fig:pbe}(b, d) presents the decomposition of the IR density change \( \Delta \rho_{\mathrm{IR}}^{\mathrm{t}} \) into the negative contribution (\( \Delta \rho_{\mathrm{IR}}^{\mathrm{t}(-)} \)) from hydrogen-bond softening, and the positive contribution (\( \Delta \rho_{\mathrm{IR}}^{\mathrm{t}(+)} \)) from interstitial accumulation. In contrast to the PBE+vdW (Fig.~\ref{fig:pbe}(b)) and PBE0+vdW (main text) case, the PBE result (Fig.~\ref{fig:pbe}(d)) shows a monotonic decrease in \( \Delta \rho_{\mathrm{IR}}^{\mathrm{t}} \) with no turnover, and \( \Delta \rho_{\mathrm{IR}}^{\mathrm{t}(+)} \) remains relatively weak across the temperature range. These findings highlight the critical role of vdW interactions in stabilizing non-H-bonded water molecules within interstitial regions. In their absence, as in the PBE case, thermal motion alone is insufficient to drive the necessary rearrangements of the H-bond network at IR that produce a density maximum. Consequently, the PBE model fails to exhibit a TMD, as shown in Table~\ref{tab:Tmelt_TMD}. These results reinforce the mechanism proposed in the main text: the density anomaly in water arises from a subtle interplay between SR and IR ordering, the latter of which depends critically on dispersion interactions.

\subsubsection*{Intermediate-Range Contribution from Disrupted Tetrahedral Structures}

To complement the analysis of IR contribution presented in the main text, we examine here the role of disrupted tetrahedral structures in contributing to the IR density variation. As shown in Fig.~\ref{fig:disrupted}(a), the oxygen–oxygen PDFs between a central water molecule adopting a disrupted tetrahedral structure and its Voronoi neighbors exhibit broader, less structured peaks compared to those of tetrahedral structures. In particular, the second-shell peak is both lower in intensity and more diffuse. Moreover, the average positions of the outer Voronoi neighbors remain nearly fixed across temperature, and the interstitial population shows little sign of inward shift. This broadened distribution indicates a more spatially dispersed packing in the disrupted tetrahedral structures, consistent with larger Voronoi cell volumes and thus lower local densities. 

The disruption of short-range (SR) ordering allows non-H-bonded water molecules to populate the interstitial region, originating from the disrupted tetrahedral structures. As a result, the ability of disrupted structures to further collapse at IR---and thereby increase the local density by \( \Delta \rho_{\mathrm{IR}}^{\mathrm{d}(+)} \)---is strongly limited (Figure~\ref{fig:disrupted}(b)), due to the rapid saturation of interstitial water molecules. As shown in Figure~\ref{fig:disrupted}(b), both the magnitude and temperature dependence of \( \Delta \rho_{\mathrm{IR}}^{\mathrm{d}} \) are significantly smaller than those associated with tetrahedral structures (see main text), suggesting that disrupted configurations play a relatively minor role in governing the density anomaly.

\subsubsection*{Mechanism Robustness on the Choice of DFT Functional Approximations}

DFT-based machine-learned potentials can exhibit quantitative deviations in absolute thermodynamic properties, including melting temperature and density. However, the present work focuses on identifying robust microscopic mechanisms underlying the density anomaly of water. To test the robustness of our findings, we investigated the density of liquid water in the temperature range around the density maximum using the same machine-learning framework, but with a deep neural network trained on data generated by SCAN functional, rather than PBE0 with van der Waals interactions as used in the main paper. In this analysis, we performed SCAN-DFT calculations and employed the DP-GEN active-learning framework to construct an accurate Deep Potential neural network. Using this SCAN-based model, we repeated the full density decomposition analysis described by Eq. (3). The results are shown in Fig.~\ref{fig:robustness}, alongside the original PBE0-vdW results for direct comparison.

Crucially, the physics observed in PBE0-vdW (Fig. R3A) is qualitatively reproduced in SCAN (Fig. R3B). Although the absolute melting temperatures and densities differ between the two functionals, the same microscopic mechanism for the density maximum emerges in both cases. Specifically, we reproduced the monotonic decrease in the density variations from changes in SR ordering, $\Delta \rho_{\text{SR}}$, and the pronounced turnover in the IR contribution, $\Delta \rho_{\text{IR}}$, which is essential for the emergence of TMD in water. We further confirmed that the turnover in $\Delta \rho_{\text{IR}}$ is mainly driven by the tetrahedral component, \( \Delta \rho_{\mathrm{IR}}^{\mathrm{t}} \), indicating that the density maximum mainly originates from H-bond structural change at IR, while the SR structure remains near-ideal tetrahedron. In contrast, the contribution from the disrupted component, \( \Delta \rho_{\mathrm{IR}}^{\mathrm{d}} \), exhibits only a weak turnover and remains relatively flat near the TMD, indicating that the H-bond network associated with locally disrupted tetrahedral ordering and a collapsed second coordination shell at IR contributes far less to the density maximum. These results demonstrate that our conclusions are not specific to a particular functional (e.g., PBE0-vdW) but reflect a general structural mechanism.

Finally, we note that the SCAN-based model predicts a TMD approximately 15 K above the melting temperature. This behavior is consistent with prior SCAN-based studies and is commonly attributed to residual self-interaction error in the SCAN functional. Such errors are expected to be significantly reduced in hybrid density functionals, including PBE0-vdW (used in the present work) and SCAN0.

\subsubsection*{Mechanism Robustness on the Choice of Structural Definitions}
The classification of local structural environments in liquid water is not unique and can depend on the choice of structural descriptors and hydrogen-bond definitions. To assess the robustness of our conclusions, we examined the sensitivity of the density decomposition to reasonable variations in both hydrogen-bond geometric criteria and coordination-number thresholds. In the main analysis, hydrogen bonds are defined using a distance criterion of $R_{\mathrm{OO}} < 3.5~\text{\AA}$ and an angular criterion of $\angle \mathrm{H{-}O\cdots O} < 30^\circ$, and tetrahedral environments are identified as configurations with four or more hydrogen bonds ($n_{\mathrm{HB}} \ge 4$). To test the robustness of this choice, we first considered an alternative hydrogen-bond definition with $R_{\mathrm{OO}} < 3.0~\text{\AA}$ and $\angle \mathrm{H{-}O\cdots O} < 35^\circ$. Fig.~\ref{fig:alter_HB} compares the decomposition of the total density change obtained using (A) the original hydrogen-bond definition and (B) the alternative definition. While the absolute population fractions of tetrahedral and disrupted environments are shifted quantitatively, the qualitative temperature dependence of the density contributions remains unchanged. In both cases, the SR contribution $\Delta \rho_{\mathrm{SR}}$ decreases monotonically with temperature, whereas the IR contribution $\Delta \rho_{\mathrm{IR}}$ exhibits a pronounced turnover near the TMD. Further decomposition shows that this turnover in $\Delta \rho_{\mathrm{IR}}$ is predominantly driven by the tetrahedral component $\Delta \rho_{\mathrm{IR}}^{\mathrm{t}}$, while the contribution from disrupted environments $\Delta \rho_{\mathrm{IR}}^{\mathrm{d}}$ remains comparatively weak and nearly temperature-independent near the TMD. This indicates that the density maximum arises from the coexistence of nearly ideal tetrahedral coordination at short range and a collapse of the hydrogen-bond network at intermediate range.

We further tested the robustness of our conclusions by modifying the coordination-number threshold used to define tetrahedral environments. Specifically, we adopted an alternative criterion in which configurations with $n_{\mathrm{HB}} \ge 3$ are classified as tetrahedral and those with $n_{\mathrm{HB}} < 3$ as disrupted. The resulting density decomposition is shown in Fig.~\ref{fig:alter_threshold}. Despite changes in the absolute population assignments, the qualitative behavior of $\Delta \rho_{\mathrm{SR}}$ and $\Delta \rho_{\mathrm{IR}}$, as well as the dominant role of tetrahedral intermediate-range contributions near the TMD, is preserved. We emphasize that defining tetrahedral coordination as configurations with four or more hydrogen bonds is physically motivated by the intrinsic tetrahedral geometry of water molecules and is widely adopted in the literature.

Taken together, these tests demonstrate that the identified microscopic mechanism underlying the density anomaly is not an artifact of a particular hydrogen-bond definition or discrete structural classification. Instead, it reflects a robust relationship between local tetrahedral ordering, intermediate-range structural rearrangements, and macroscopic density variations in liquid water.

\newpage


\begin{figure}[h]
\centering
\includegraphics[width=0.6\linewidth]{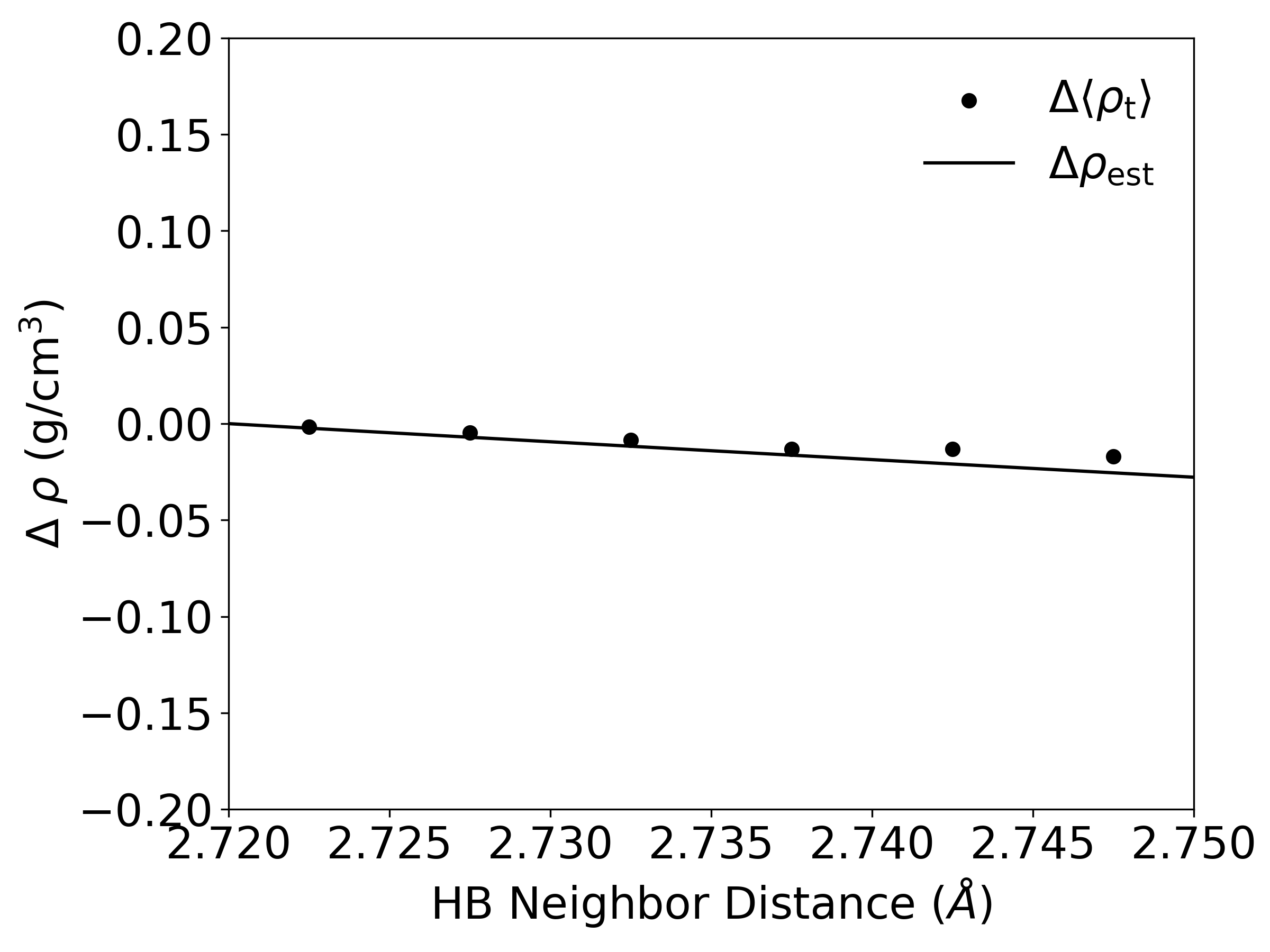}
\caption{
\textbf{Comparison between the actual and estimated mean local densities of water molecules adopting tetrahedral structures, as a function of the average H-bond neighbor distance.}
}
\label{fig:HBlength}
\end{figure}

\newpage

\begin{figure}[h]
\centering
\includegraphics[width=0.6\linewidth]{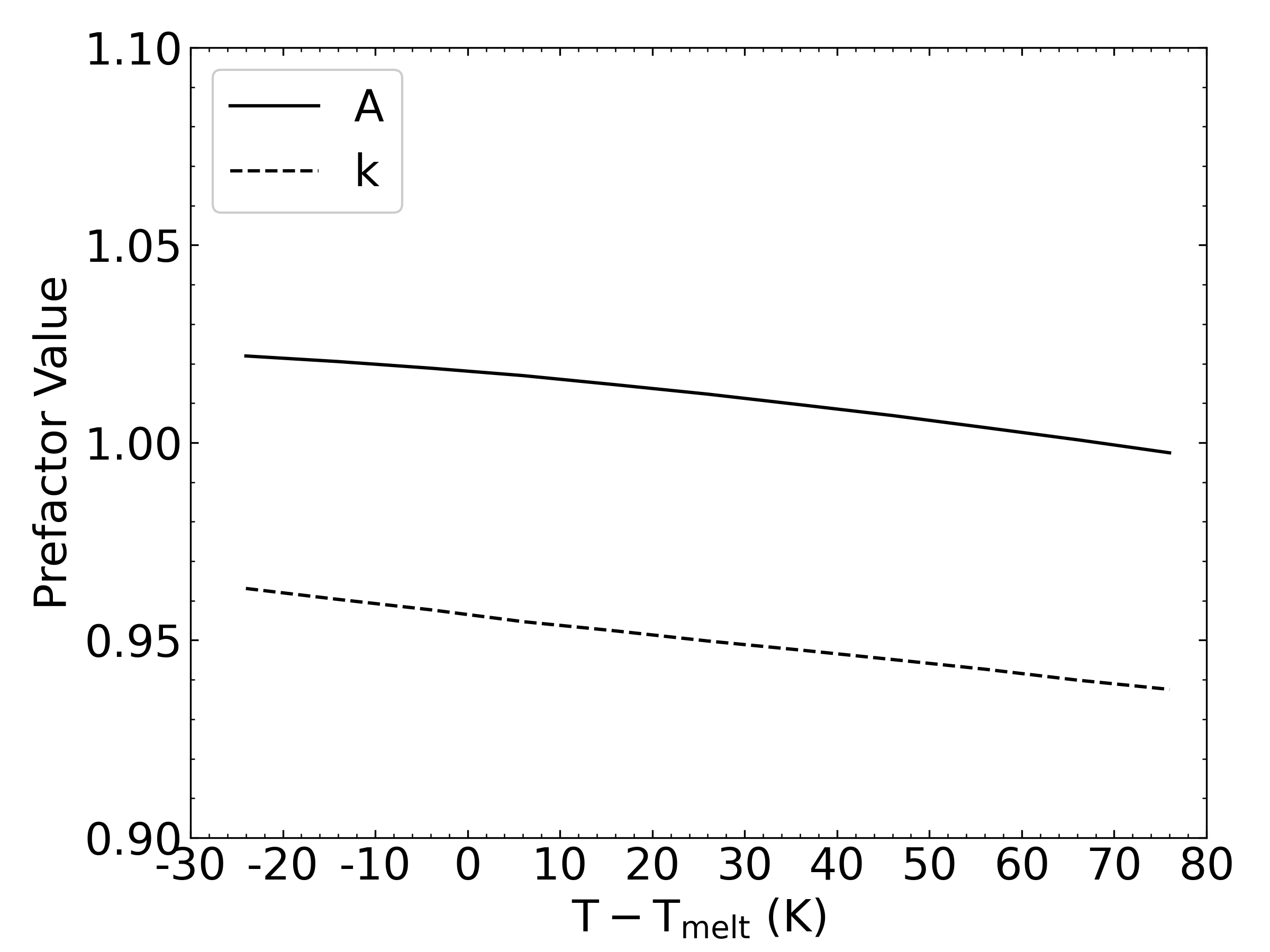}
\caption{
\textbf{Temperature dependence of the prefactors $A$ and $k$.}
}
\label{fig:prefactors}
\end{figure}

\newpage

\begin{figure}[h]
\centering
\includegraphics[width=0.6\linewidth]{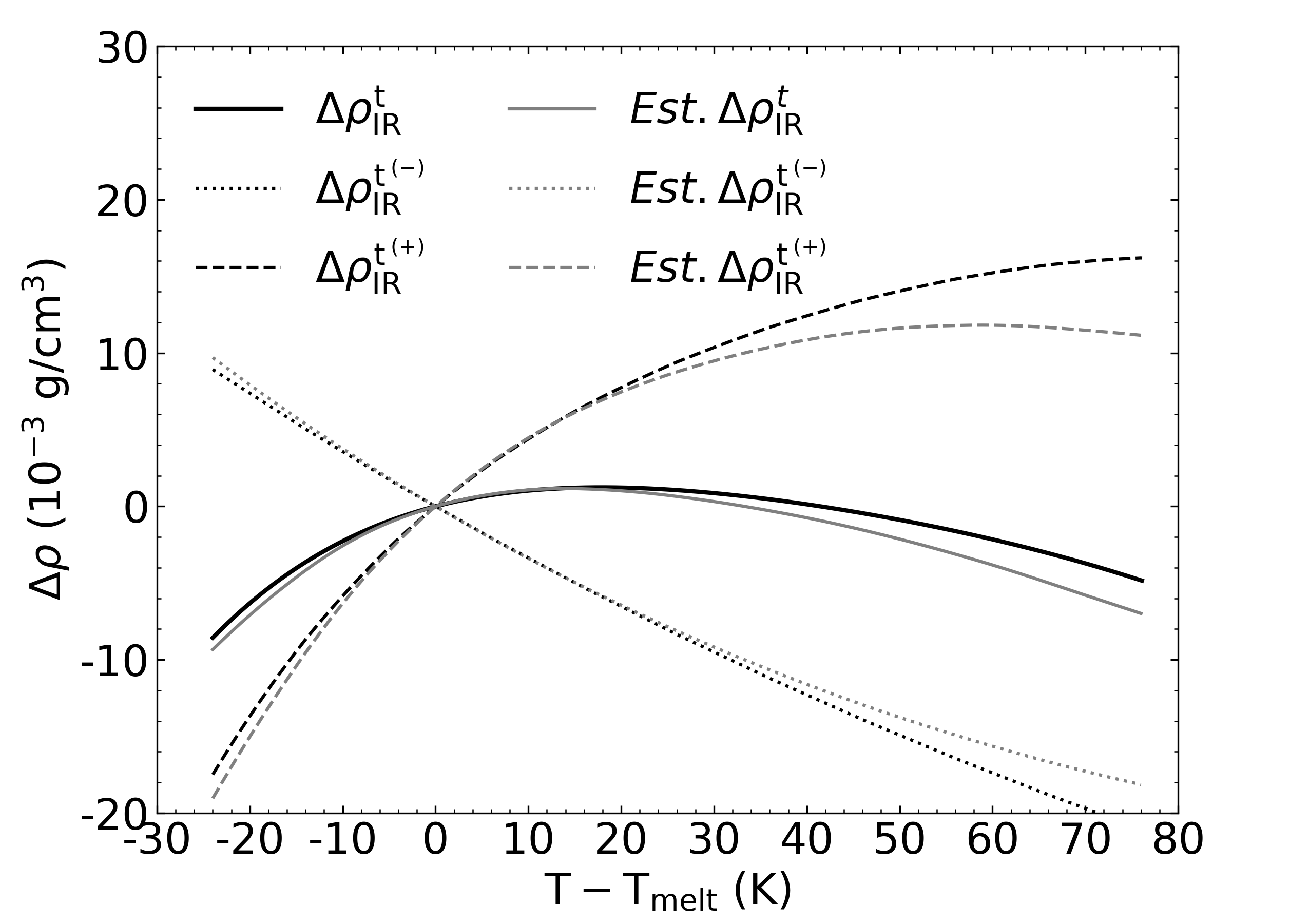}
\caption{
\textbf{Quantitative estimate of the impact of treating $A$ and $k$ as constants on the calculated density changes.}
}
\label{fig:prefactor_comparison}
\end{figure}

\newpage

\begin{figure}[h]
\centering
\includegraphics[width=0.85\linewidth]{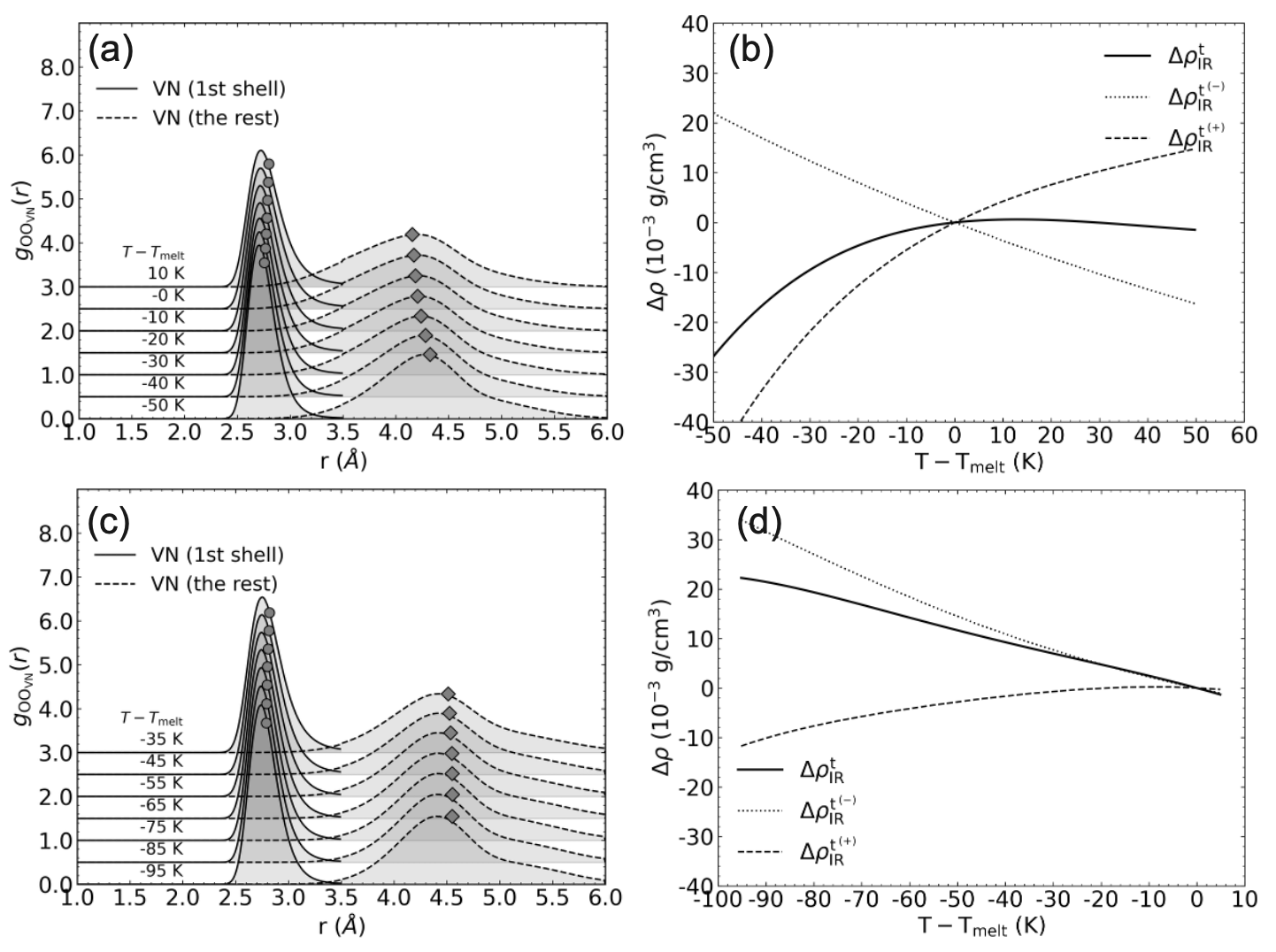}
\caption{
\textbf{Intermediate-range contributions with and without vdW interactions.} Results obtained using PBE+vdW functional (a, b) and PBE functional (c, d).
\textbf{(a, c)} Temperature dependent PDFs between a central water molecule forming a tetrahedral structure and its Voronoi neighbors. Each PDF is decomposed into contributions from Voronoi neighbors in the first coordination shell and those beyond it, with circles and diamonds indicating the average radial positions of each group, respectively.
\textbf{(b, d)} Decomposition of IR density change of disrupted tetrahedral structures \( \Delta \rho_{\mathrm{IR}}^{\mathrm{t}} \) into \(\Delta \rho_{\mathrm{IR}}^{\mathrm{t}\,^{(-)}}\) and \(\Delta \rho_{\mathrm{IR}}^{\mathrm{t}\,^{(+)}}\).
}
\label{fig:pbe}
\end{figure}

\newpage

\begin{figure}[h]
\centering
\includegraphics[width=0.85\linewidth]{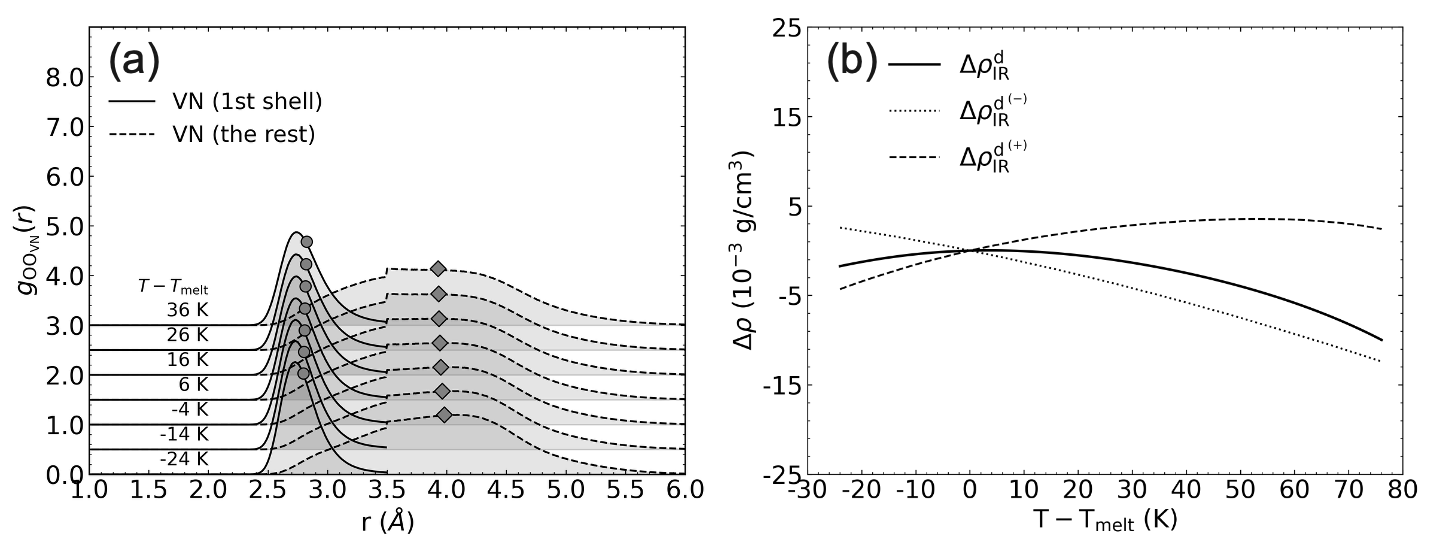}
\caption{\textbf{Intermediate-range contributions of disrupted tetrahedral structures.}
\textbf{(a)} Temperature dependent PDFs between a central water molecule forming a disrupted tetrahedral structure and its Voronoi neighbors. Each PDF is decomposed into contributions from Voronoi neighbors in the first coordination shell and those beyond it, with circles and diamonds indicating the average radial positions of each group, respectively.
\textbf{(b)} Decomposition of IR density change of disrupted tetrahedral structures \( \Delta \rho_{\mathrm{IR}}^{\mathrm{d}} \) into \(\Delta \rho_{\mathrm{IR}}^{\mathrm{d}\,^{(-)}}\) and \(\Delta \rho_{\mathrm{IR}}^{\mathrm{d}\,^{(+)}}\).
}
\label{fig:disrupted}
\end{figure}

\begin{figure}[h]
\centering
\includegraphics[width=0.85\linewidth]{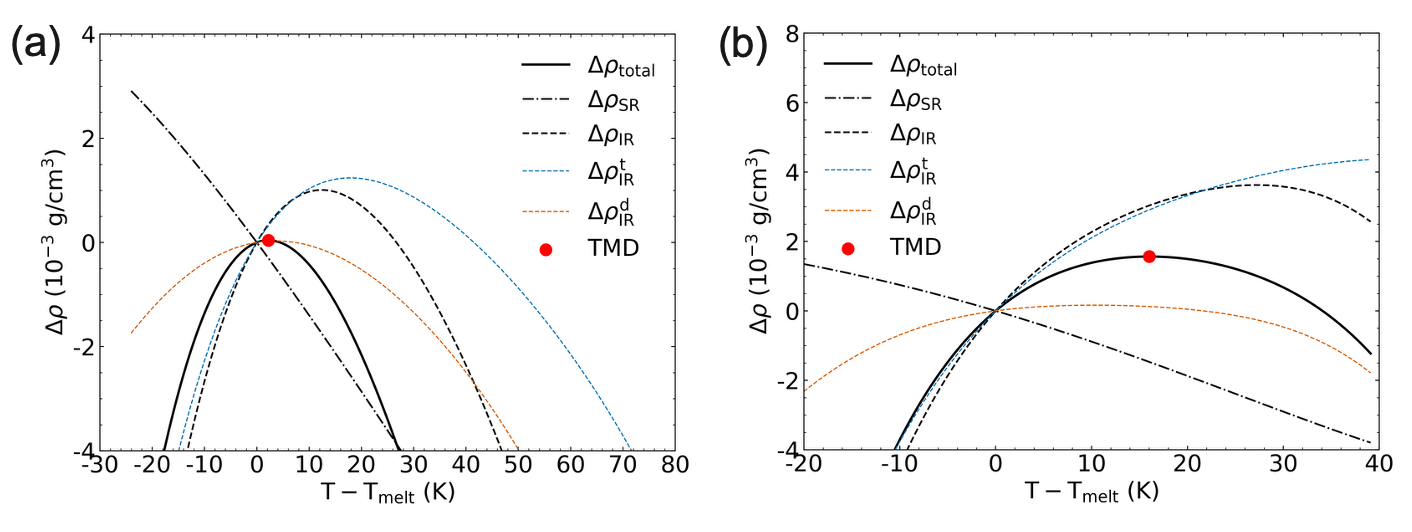}
\caption{\textbf{Decomposition of the total density change using (A) PBE0-vdw and (B) SCAN functional.}
}
\label{fig:robustness}
\end{figure}

\begin{figure}[h]
\centering
\includegraphics[width=0.85\linewidth]{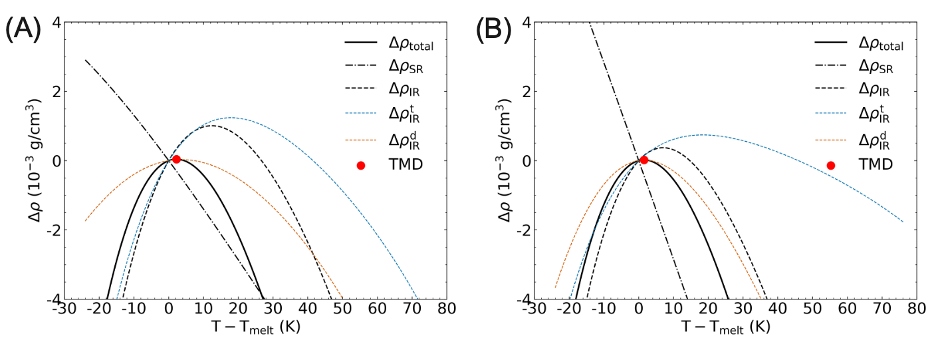}
\caption{\textbf{Decomposition of the total density change using (A) the original HB definition ($R_{\mathrm{OO}} < 3.5$~\AA\, $\angle \mathrm{H_D{-}O_D{\cdots}O_A} < 30^\circ$) and (B) an alternative criterion ($R_{\mathrm{OO}} < 3.0$~\AA\, $\angle \mathrm{H_D{-}O_D{\cdots}O_A} < 35^\circ$).}
}
\label{fig:alter_HB}
\end{figure}

\begin{figure}[h]
\centering
\includegraphics[width=0.85\linewidth]{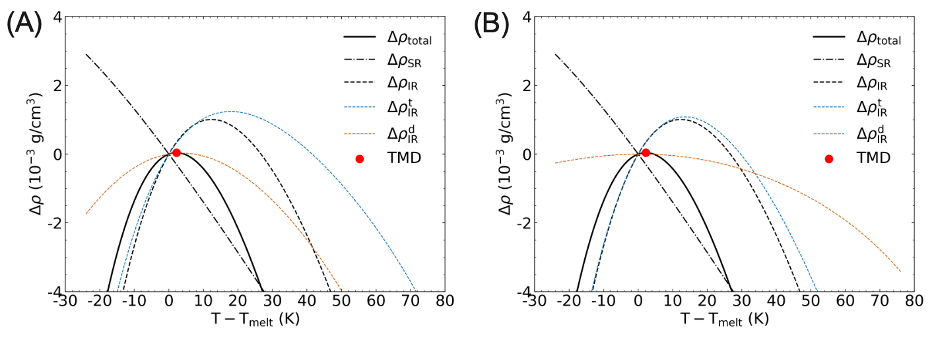}
\caption{\textbf{Decomposition of the total density change using (A) the original (“tetrahedral” $n_{\mathrm{HB}} \ge 4$, “disrupted tetrahedral” $n_{\mathrm{HB}} < 4$) and (B) the alternative criterion (“tetrahedral” $n_{\mathrm{HB}} \ge 3$, “disrupted tetrahedral” $n_{\mathrm{HB}} < 3$).}
}
\label{fig:alter_threshold}
\end{figure}

\newpage




\begin{table}[h]
\centering
\caption{\textbf{Mean radial positions ($r$) and standard deviation ($\sigma$) for Voronoi neighbors in the first coordination shell
and beyond at different temperatures (referencing Fig.~3A).}}
\begin{tabular}{l c c c c}
\hline
& \multicolumn{2}{c}{First Shell} & \multicolumn{2}{c}{The Rest} \\
Temperature (K) & $r$ (Å) & $\sigma$ (Å) & $r$ (Å) & $\sigma$ (Å) \\
\hline
290 & 2.782 & 0.161 & 4.229 & 0.534 \\
300 & 2.789 & 0.167 & 4.200 & 0.539 \\
310 & 2.796 & 0.173 & 4.177 & 0.543 \\
320 & 2.802 & 0.178 & 4.159 & 0.548 \\
330 & 2.809 & 0.183 & 4.144 & 0.553 \\
340 & 2.815 & 0.187 & 4.135 & 0.559 \\
350 & 2.821 & 0.191 & 4.124 & 0.564 \\
360 & 2.826 & 0.195 & 4.117 & 0.569 \\
370 & 2.832 & 0.199 & 4.111 & 0.574 \\
380 & 2.837 & 0.202 & 4.107 & 0.580 \\
390 & 2.843 & 0.206 & 4.105 & 0.586 \\
\hline
\end{tabular}
\label{tab:variance}
\end{table}

\newpage

\begin{table}[h]
\centering
\caption{\textbf{AIMD trajectory lengths at different temperatures for PBE, PBE+vdW, and PBE0+vdW functionals.}}
\begin{tabular}{lcccccccccc}
\hline
Temperature (K) & 290 & 305 & 335 & 365 & 395 \\
PBE (ps) & 43.3 & 44.6 & 43.8 & 41.4 & 43.5 \\
\hline
Temperature (K) & 260 & 275 & 290 & 305 & 320 & 335 & 350 & 365 & 380 & 395 \\
PBE+vdW (ps) & 13.5 & 13.4 & 20.2 & 60.2 & 56.3 & 61.6 & 13.5 & 61.2 & 13.7 & 58.7 \\
\hline
Temperature (K) & 300 & 330 & 360 & 390 \\
PBE0+vdW (ps) & 13.0 & 12.8 & 11.2 & 11.0 \\
\hline
\end{tabular}
\label{tab:aimd_trajectories}
\end{table}

\newpage

\begin{table}[h]
\centering
\caption{\textbf{Fitted fourth-order polynomial coefficients ($a_0$--$a_4$) used for TMD calculation. The polynomial is given by $\rho(T) = a_0 + a_1 T + a_2 T^2 + a_3 T^3 + a_4 T^4$, with $T$ in K and $\rho$ in g/cm$^3$.}}
\label{tab:polyfit_tmd}
\begin{tabular}{ccccc}
\hline
$a_0$ & $a_1$ & $a_2$ & $a_3$ & $a_4$ \\
\hline
-7.54187 & 9.46856$\times 10^{-2}$ & -3.92944$\times 10^{-4}$ & 7.27580$\times 10^{-7}$ & -5.09533$\times 10^{-10}$ \\
\hline
\end{tabular}
\end{table}

\newpage

\begin{table}[h]
\centering
\caption{\textbf{Melting temperatures ($T_\mathrm{melt}$) and temperatures of maximum density (TMD) for experiment and different exchange-correlation functionals.}}
\begin{tabular}{l c c c c}
\hline
Temperature (K) & Experiment & PBE & PBE+vdW & PBE0+vdW \\
\hline
$T_\mathrm{melt}$ & 273.15 & 385.2 & 340.3 & 314.0 \\
TMD & 277.13 & --- & 337.4 & 315.6 \\
\hline
\end{tabular}
\label{tab:Tmelt_TMD}
\end{table}

\newpage




\clearpage 





\end{document}